# Autoencoder-based integrative multi-omics data embedding that allows for confounder adjustments

Tianwei Yu[1]

[1] School of Data Science, The Chinese University of Hong Kong – Shenzhen, Shenzhen, Guangdong Province, China 518172. Email: yutianwei@cuhk.edu.cn

## Abstract

In the integrative analyses of omics data, it is often of interest to extract data representation from one data type that best reflect its relations with another data type. This task is traditionally fulfilled by linear methods such as canonical correlation analysis (CCA) and partial least squares (PLS). However, information contained in one data type pertaining to the other data type may be complex and in nonlinear form. Deep learning provides a convenient alternative to extract low-dimensional nonlinear data embedding. In addition, the deep learning setup can naturally incorporate the effects of clinical confounding factors into the integrative analysis. Here we report a deep learning setup, named Autoencoder-based Integrative Multi-omics data Embedding (AIME), to extract data representation for omics data integrative analysis. The method can adjust for confounder variables, achieve informative data embedding, rank features in terms of their contributions, and find pairs of features from the two data types that are related to each other through the data embedding. In simulation studies, the method was highly effective in the extraction of major contributing features between data types. Using a real microRNA-gene expression dataset, and a real DNA methylation-gene expression dataset, we show that AIME excluded the influence of confounders including batch effects, and extracted biologically plausible novel information. The R package based on Keras and the TensorFlow backend is available at https://github.com/tianwei-yu/AIME.

## 1 Introduction

In more and more studies, multiple omics data are collected on the same set of subjects to obtain a global view of the molecular signature of a disease. When analyzing such data, a common task is to find data embedding in a lower dimensional space from one data type that best preserves the information pertaining to another data type. Such data embedding can reveal mechanistic relations between the data types, or serve as extracted predictors in predictive models.

To achieve data embedding while considering two data types, the most common methods are dimension reduction approaches including Canonical Correlation Analysis (CCA), Partial Least Squares (PLS) and their variants, which are based on (sparse) linear projections of the data [1-3]. Given the complexity of omics data, nonlinear equivalents to such linear methods were developed, such as kernel-based [4] and deep learning-based CCA [5]. Beside dimension reduction, factorization and clustering techniques were also developed to analyze multiple data types collected on the same set of samples [6, 7]. In biomedical studies, it is often of interest to extract data representation while adjusting for linear/non-linear effects of clinical confounding factors, such as age, gender, ethnicity, batch effects *etc*. Current approaches do not adjust for confounders, which could lead to data embedding dominated by the confounding factors, masking the true biological relations between the data types.

Autoencoder is a deep learning – based nonlinear embedding approach that is typically used to achieve sparse data representation from a single dataset [8], reduce noise [9], impute missing values [10], conduct pre-training for classification tasks [11], and make functional inferences [12-15]. A supervised auto-encoder (SAE) is an auto-encoder with the additional supervised loss component on the representation layer, which factors in a second data type, with a goal of improving generalization performance [16]. Variants of autoencoders have been used in combining multiple data matrices. In terms of integrative analysis, joint learning schemes [17, 18] are used to combine multi-omics data in both the input and reconstruction layers in order to find their interactions for the prediction task. Combining joint learning with probabilistic Gaussian Mixture Model was able to learn informative joint latent features to construct the association between omics data and find cell heterogeneity [19].

In this study, our goal is geared towards data interpretation. We aim at achieving nonlinear data embedding from one omics data type, in order to preserve the information pertaining to another omics data type. Prediction or reconstruction is not a major concern. We modify the autoencoder structure by using two data types in the input and reconstruction layers respectively, and allowing contributions from clinical confounding factors by including them as auxiliary inputs at the representation layer. The approach is different than existing joint learning methods [17-19] in that it doesn't seek joint embedding which could be a mixture of the two data types, in which their individual contributions are hard to separate, and it doesn't involve outcome variables. It is also different than Deep learning-based

CCA [5], which tries to find two separate embeddings from the two data types that are highly correlated, and doesn't allow the adjustment of confounders. Here our main interest is to find features in one data type that influence the other data type.

The method is named Autoencoder-based Integrative Multi-omics data Embedding (AIME). In simulations, we show the method can effectively extract influential features. When sample size is large, the method can be more sensitive than CCA and PLS even when all relations are linear. In real data analysis, the method can exclude the superficial relations caused by clinical confounding factors, and extract meaningful miRNAs and methylation sites that influence gene expression.

## 2 Methods

### 2.1. The setup

Assume there are two types of high-throughput measurements on the same set of samples. Let $\boldsymbol{X}_{\boldsymbol{N}\times\boldsymbol{p}}$ denote the first data type, where there are $\boldsymbol{p}$ features and $\boldsymbol{N}$ samples. Let $\boldsymbol{Y}_{\boldsymbol{N}\times\boldsymbol{q}}$ denote the second data type, where there are $\boldsymbol{q}$ features and $\boldsymbol{N}$ samples. Our interest is to extract a low-dimensional nonlinear data embedding from the $\boldsymbol{X}$ matrix, $\boldsymbol{E}_{\boldsymbol{N}\times\boldsymbol{r}}$, where $\boldsymbol{r}$ is small, such that the $\boldsymbol{E}$ matrix contains as much information to nonlinearly reconstruct the $\boldsymbol{Y}$ matrix as possible.

We set up a neural network structure that is similar to autoencoder, as shown in Fig. 1. Different from the typical autoencoder, the input layer and reconstruction layer use two different omics data types. The input layer contains $\boldsymbol{p}$ variables corresponding to the columns of the $\boldsymbol{X}$ matrix. The output layer contains $\boldsymbol{q}$ variables corresponding to the columns of the $\boldsymbol{Y}$ matrix. In addition to the input data $\boldsymbol{X}$, clinical confounders such as age, gender, ethnicity, batch *etc*, can form another matrix $\boldsymbol{C}_{\boldsymbol{N}\times\boldsymbol{s}}$, Their effects can be adjusted for by inserting the variables in $\boldsymbol{C}$ as auxiliary variables at the bottleneck layer. This encourages the model to find nonlinear data embedding of $\boldsymbol{X}$ that contribute to the reconstruction of $\boldsymbol{Y}$, independent of the clinical confounders.

In some sense this is a prediction structure with very high dimensional outcome. Such a prediction task is unrealistic, and our goal is not prediction. With a very narrow bottleneck layer in the middle, we essentially seek a nonlinear dimension reduction of the input data $\boldsymbol{X}$, which best preserves the information pertaining to the output data ***Y**,* while adjusting for confounding factors in matrix ***C***. Following traditional statistical terminology in dimension reduction, we call the columns of the embedded data matrix ***E*** "components" in this manuscript.

## 2.2. Implementation

The program was implemented in R using the Keras neural networks API [20], to facilitate users of R to conduct the analysis. The implementation requires both R and the TensorFlow backend. With regard to the sizes of the layers of the network, the method allows three different ways for the user to specify. (1) The user can directly specify the sizes of all the individual layers; (2) the user can input a shrinkage factor, such that the size of each layer in the encoder is the product of the size of the previous layer and the shrinkage factor, and the size of each decoder layer is the product of the next layer and the shrinkage factor; (3) the user can input the desired number of input/out layers, and the shrinkage factor is calculated based on the number of layers. Dropout rates can be specified to be uniform across all layers, or in a layer-by-layer manner.

Given the number of layers and dropout rates, the data is split into training and testing sets. The prediction error rate on the testing set is used to select the number of training epochs. Once the number of epochs is determined, the full dataset is used to fit the model again.

## 2.3. Estimating feature importance

To find which feature from the input matrix ***X*** is more influential, we use a permutation scheme. We fix the parameters in the trained model. In each iteration, one variable in the ***X*** matrix is permuted, and new embedding is calculated based on the existing parameters. We then compare the new embedded data with the embedded data from the unpermuted data. The amount of location shift, measured by the sum of squared distances across all the embedded data points, is taken as the importance of the permuted variable. Similarly, we estimate the pairwise influence, *i.e.* the influence of one variable in the ***X*** matrix on one variable in the ***Y*** matrix in the same permutation, by recording the amount of change of each ***Y*** variable, when the ***X*** variables are permuted.

## 2.4. Tuning hyperparameters

In this study, we used the multivariate skewness and kurtosis of the embedded data to select the number of layers and dropout rates. At each hyperparameter setting, the data embedding (matrix $\boldsymbol{E}$) is computed, and the average absolute pairwise correlation between the columns of the $\boldsymbol{E}$ matrix is calculated. Among the settings for which the average correlation is below a threshold, the Mardia's multivariate skewness and kurtosis coefficients are calculated for the embedded data [21]. We rank each setting by the skewness and kurtosis of the embedded data, and then select the setting the yield the highest average rank of skewness and kurtosis. This process selects parameter settings that yield embedding that is not highly correlated, as well as with a distribution far from multivariate normal. This is because a random projection of the data into lower dimensions tend to yield multivariate normal distribution. The criterion is similar to that of projection pursuit [22].

2.5. Simulation study

We use the following procedure to generate simulated data：

(1) Generate the $\boldsymbol{X}$ matrix with $n_x$ variables and $N$ samples using multivariate normal distribution. The mean vector is $\boldsymbol{0}$. The diagonal elements of the variance-covariance matrix $\Sigma$ is 1, and all off-diagonal elements take value $\rho$, which is a value between 0 and 1. Inverse-normal transform $X$, and subtract 0.5, such that the values in the $X$ matrix are between -0.5 and 0.5.
(2) Select the first $k$ variables in the $\boldsymbol{X}$ matrix. Generate three linear combinations of the $k$ variables $z_m = \sum_{j=1}^{k} \beta_{mj} x_j, m = 1,2,3$, where the $\beta's$ are randomly sampled from a uniform distribution on $[-2,-1] \cup [1,2]$.
(3) Generate the first $m \times k$ variables in the $Y$ matrix, by first generating a linear combination of the $z$ variables, $r_j = \alpha_{j1} z_1 + \alpha_{j2} z_2 + \alpha_{j3} z_3, j = 1, \ldots, mk$, where the $\alpha's$ are randomly sampled from a uniform distribution on $[-2,-1] \cup [1,2]$. Then the $r$ variables are re-scaled to facilitate nonlinear transformation, by subtracting the mean and dividing by 3 times the standard deviation. This transformation ensures most of the $r$ values are between -1 and 1. We then take $y_j = f_j(r_j), j = 1, \ldots, mk$, where $f_j()$ is sampled from five different function: (1): $f(r)=r$, (2): $f(r) = |r|$, (3): $f(r) = sin((5 \times (r + 0.5) \times pi))$, (4): $f(r) = (2r)^2$, and (5) a step function that takes value 1 when $r$ is between the $25^{th}$ and $75^{th}$ *percentiles*, and 0 otherwise. The proportion of $y$'s that receive the original $r$ values without any transformation is controlled by a hyperparameter. Gaussian random noise is added to each of the $y$ variables, such that the noise variance is 1/10 of that of the $y$ variable.

(4) The remaining $n_y - mk$ variables in the $Y$ matrix are sampled from multivariate normal distribution of mean vector is $\boldsymbol{0}$, and variance-covariance matrix with diagonal value 1 and off-diagonal value $\rho$. All the variables in the $Y$ matrix are then re-scaled to have mean 0 and standard deviation 1.

We then analyze the simulated data using three methods: AIME, CCA and PLS. For simplicity, we fix the number of input layers and output layers of AIME at 3, and the dropout rate at 0.4. We compare the variable importance ranking generated by each method, *i.e.* whether the first $k$ variables in the $X$ matrix receive higher importance scores, using the area under the curve (AUC) of the precision-recall (PR) curve.

A number of scenarios, as specified by the combinations of $n_x, n_y, k, m, N, \rho$, and the proportion of *y's* that are linearly associated with $\boldsymbol{X}$, *i.e.* receiving the original *r* values without any nonlinear transformation. In each scenario, the simulation was repeated 10 times, and the average PR-AUC value was taken.

## 3 Results and Discussion

### 3.1 Simulation Results

The simulation results are shown in Figure 2. Because most of the simulated ***X*** variables do not contribute to the relation with ***Y***, we used the area under the precision-recall curve (PR-AUC) to assess the methods' capability to separate the truly contributing ***X*** variables from the rest.

When the relation between ***X*** and ***Y*** were purely non-linear (Figure 2, left column), only AIME could extract the contributing ***X*** variables. However, it required the sample size to be relatively large when the number of impacted ***Y*** variable were small. For example, when 10 ***X*** variables contributed to the relation between ***X*** and ***Y***, and a total of 100 ***Y*** variables (out of 1000) were impacted, AIME could only detect some signal when the sample size was 10,000. When a moderate number of ***Y*** variables were associated with the ***X*** variables (400 out of 1000; Figure 2, second plot in left column), the power to select the contributing ***X*** variables became higher. Interestingly, we observed that the power was higher when there was a moderate level of correlations (0.3) among all the ***X*** variables, as compared with the scenarios of no correlation or higher correlation (0.6). Further increasing the number of impacted ***Y*** variables helped improve the power the detect the contributing ***X*** variables to some extent. Due to the intrinsic difficulty in capturing nonlinear relations, usually higher sample sizes (>1000) was required to achieve moderate PR-AUC levels.

We next examined the situation where the relation between ***X*** and ***Y*** were mixed (Figure 2, center column). When the number of contributing X and Y variables were low (Figure 2, center column, top plot), AIME trailed behind PLS and CCA in the capability to detect contributing ***X*** variables. When the number of contributing ***X*** variables increased to 20, and 40% of ***Y*** variables were influenced by them, AIME achieved similar level of performance as PLS and CCA. Further increasing the number of contributing ***X*** variables or the number of ***Y*** variables impacted by them, AIME surpassed the other two methods (Figure 2, center column, lower plots). We note this was achieved at very low sample size of 200, indicating the capability of neural networks even in the situation of $N<<p$, i.e. sample size much smaller than variable count.

When all the relations between ***X*** and ***Y*** were linear (Figure 2, right column), the relative performance of the methods were similar to that of the mixed situation (Figure 2, center column). Again we saw AIME trailing PLS and CCA when the number of contributing X variables were small (Figure 2, right column, top plot), but rising to a better performance when the number of contributing ***X*** variables increased, and surpassing PLS and CCA when the number of impacted ***Y*** variables were high, even at very small sample size (Figure 2, right column, lower plots).

### 3.2 TCGA BRCA microRNA and gene expression dataset

We analyzed the TCGA microRNA and gene expression datasets [23]. After log-transforming both data matrices, we matched the common subjects in which both data were measured. We then filtered the microRNA data by selecting microRNAs with a coefficient of variation (CV) larger than 0.25, and filtered the gene expression data by selecting genes with a CV larger than 0.2. The resulting matrix dimensions were 451×242 for the microRNA data, and 451×6086 for the gene expression data.

We then used the microRNA data as input, and the gene expression data as output. First we ran the analysis without adjusting for any confounder. The method selected 4 layers for the encoder and 5 layers for the decoder. The resulting embedded data is shown in Figure 3, upper-right triangle. We can see that clearly the embedded data were separated based on the PAM50 (Prosigna Breast Cancer Prognostic Gene Signature Assay) subtypes, which is based on a multi-gene signature for risk stratification [24]. As the ER (estrogen receptor) status is highly correlated with the PAM50 score, the embedded data also separated the subjects based on ER status very well (Supporting Figure 1).

We then adjusted for age, T1 (tumor size) status, and ER status in the analysis. The resulting embedded data clearly lost the relation with PAM50 subtypes, with all four PAM50 subtypes mixed together (Figure 3, subplots in the lower-left triangle). Similar effect was observed with ER status (Supporting Figure 1, subplots in the lower-left triangle).

We then examined the functionality of the major contributing microRNAs by selecting the top 20 microRNAs based on the importance score and applying DIANA-miRPath v3.0 [25]. Table 1 shows the significant pathways over-represented by the top microRNAs with (Table 1, left column) and without (Table 1, right column) adjusting for age, T1 status, and ER status. In both scenarios with/without confounder adjustments, we found most of the significant pathways belonged to 3 major categories – cancer, signal transduction, and neuronal functions. With adjustment to the confounders, more pathways became significant, including the critically important pathway in breast cancer – estrogen signaling pathway.

We further examined which genes were mostly impacted by the top microRNAs at the expression level. Using results from the analysis that adjusted for age, T1 status, and ER status, we took the microRNA-gene importance, and selected the columns corresponding to the top 20 microRNAs. We normalized each column (each corresponding to on microRNA) to the same overall weight, and selected the microRNA-gene pairs with weights within the top 1%. Figure 4 shows the selected microRNA-gene pairs. We observed that some microRNAs impacted many genes in the top 1% gene list, while some impacted relatively few in this list, indicating their influences on gene expression data were more pervasive and less focused. Functional analysis using GOstats [26] of the selected genes (Table 2) showed the genes were highly concentrated in three categories of biological functions – blood vessel development/hemopoiesis, cell migration and adhesion, and immune cell functions. The results indicate that after adjusting for age, T1 status, and ER status, the association between microRNA expression and gene expression are highly focused on the biological functions associated with solid tumor development and immune cell infiltration.

### 3.3 GSE56046/GSE56047 DNA methylation and gene expression dataset

We analyzed the GSE56046/GSE56047 dataset from the MESA methylomics and transcriptomics study, which was conducted on purified human monocytes from a large study population [27]. We filtered the normalized gene expression data with the criterion of coefficient of variation (CV) > 0.05, and the DNA methylation M-values with the criterion of standard deviation > 1.25. The resulting data contained 5459 genes and 5703 CpG loci, both measured in the same 1202 subjects.

We focused on the nonlinear embedding of methylation data that was associated with gene expression. The M-value matrix was used as input of the encoder, and the gene expression matrix was used as output. Figure 5 shows the embedded data in four dimensions by AIME without adjusting for the "racegendersite" variable of the dataset, which is a combined factor coding for the combinations of race, gender, and data collection site (Figure 5, subplots in upper right triangle). As is clearly demonstrated by the separation of the colored data points, the AIME results indicate the association between DNA methylation and gene expression is strongly dependent on the study site, race and gender. We tested all the other available factors in the data, and no other clear relation to the embedded data pattern was found. Reversing the roles of the methylation data and the gene expression data, *i.e.* using gene expression data as input, the pattern of embedded data showed an even stronger association with the "racegendersite" variable (Supporting Figure 2, upper-right triangle subplots).

We then conducted the analysis by adjusting for age and the "racegendersite" variable. The influence of age, gender and study site was fully removed from the embedding when methylation data was used as input (Figure 5, subplots in lower-left triangle), while small portion of the influence remained when gene expression data was used as input (Supporting Figure 2, subplots in lower-left triangle). Based on the loci annotations provided in the GSE56046/GSE56047 dataset, we analyzed the biological processes associated with the top methylation loci that showed the largest impact on gene expressions. Based on the distribution of the importance scores, for unadjusted analysis, we selected the top 447 loci, which were then annotated to 247 genes. Using the GOstats package, 21 biological processes were significant at the p-value threshold of 0.01. For adjusted analysis, we selected the top 376 loci, which were annotated to 205 genes. Thirty-four biological processes were significantly over-represented at the 0.01 level. In both sets of significant biological processes, the majority were immune-system processes. This agrees very well with the fact that the data were generated from human monocytes.

The adjustment to the confounder variables resulted in more significant processes, as well as more focused and interpretable biological processes (Table 3).

### 3.4. Discussions

AIME can be seen as a nonlinear equivalent to CCA, with the added capability to adjust for confounder variables. Besides being able to extract nonlinear relationships that traditional methods cannot, when sample size is large enough, AIME is even more effective than traditional linear methods such as CCA and PLS in extracting linear relationships. In real data applications, AIME was able to exclude the influence of unwanted confounders and extract novel patterns. The results were easily interpretable. We believe AIME is a valuable addition to the current methods of omics data integrative analyses.

## Acknowledgements

The author thanks Dr. Yunchuan Kong and Dr. Hao Wu for helpful discussions. This work has been partially supported by Shenzhen Research Institute of Big Data (SRIBD) and the University Development Fund of CUHK-Shenzhen.

Table 1. Functional analysis of the important miRNAs using mirPath 3.0[#].

| **No confounders** | **With confounders** |
|---|---|
| Prion diseases | Prion diseases |
| FoxO signaling pathway | FoxO signaling pathway |
| TGF-beta signaling pathway | ***Estrogen signaling pathway*** |
| ***Proteoglycans in cancer*** | TGF-beta signaling pathway |
| Axon guidance | ***Lysine degradation*** |
| Endocytosis | ***GABAergic synapse*** |
| ***Signaling pathways regulating pluripotency of stem cells*** | ***Thyroid hormone synthesis*** |
| ***Hippo signaling pathway*** | Axon guidance |
| Thyroid hormone signaling pathway | Endocytosis |
| ***Regulation of actin cytoskeleton*** | ***Mucin type O-Glycan biosynthesis*** |
| ***Glutamatergic synapse*** | Pancreatic cancer |
| Renal cell carcinoma | ***Biotin metabolism*** |
| Pancreatic cancer | ***cGMP-PKG signaling pathway*** |
| Morphine addiction | ***Prolactin signaling pathway*** |
| | Morphine addiction |
| | ***Prostate cancer*** |
| | Renal cell carcinoma |
| | Thyroid hormone signaling pathway |

[#] Pathways with a p-value <0.001 are listed. Bold and underlined pathways are those different between the two lists.

**Table 2.** Top biological pathways associated with the top 1% of genes related to the 20 microRNAs, after adjusting for age, T1 (tumor size) status, and estrogen receptor (ER) status[#].

| GOBPID | Pvalue | Term |
|---|---|---|
| GO:0072359 | 2.71E-09 | circulatory system development |
| GO:0030097 | 6.42E-08 | hemopoiesis |
| GO:0030198 | 1.60E-07 | extracellular matrix organization |
| GO:0002521 | 1.67E-07 | leukocyte differentiation |
| GO:0030334 | 4.70E-07 | regulation of cell migration |
| GO:0098609 | 6.88E-05 | cell-cell adhesion |
| GO:0071363 | 7.47E-05 | cellular response to growth factor stimulus |
| GO:0050863 | 9.24E-05 | regulation of T cell activation |
| GO:0071345 | 9.24E-05 | cellular response to cytokine stimulus |
| GO:0080134 | 9.32E-05 | regulation of response to stress |

[#]Some highly overlapping processes were manually removed.

**Table 3.** Top biological pathways associated with the important M-values[#].

| GOBPID | Pvalue | Term |
|---|---|---|
| AIME unadjusted (21 total) | | |
| GO:0002250 | 0.000135 | adaptive immune response |
| GO:0071346 | 0.00241 | cellular response to interferon-gamma |
| GO:0061061 | 0.00372 | muscle structure development |
| GO:0031344 | 0.00653 | regulation of cell projection organization |
| GO:0002495 | 0.00654 | antigen processing and presentation of peptide antigen via MHC class II |
| AIME adjusted (34 total) | | |
| GO:0002250 | 2.34E-06 | adaptive immune response |
| GO:0051304 | 3.50E-05 | chromosome separation |
| GO:0071346 | 0.000104 | cellular response to interferon-gamma |
| GO:0002449 | 0.000173 | lymphocyte mediated immunity |
| GO:0050776 | 0.00373 | regulation of immune response |

[#]Some highly overlapping processes were manually removed.

.

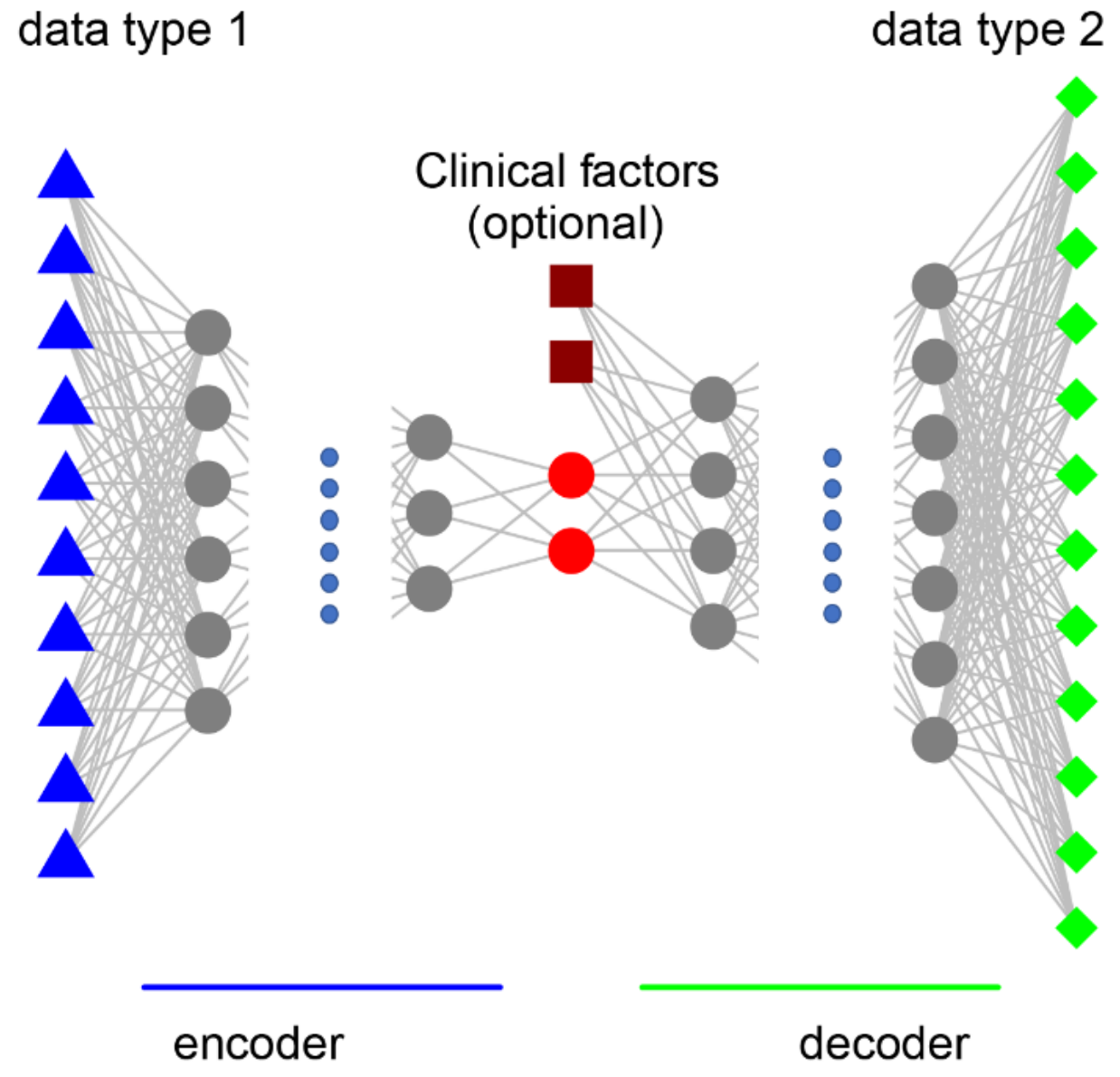

**Figure 1.** The setup of the model.

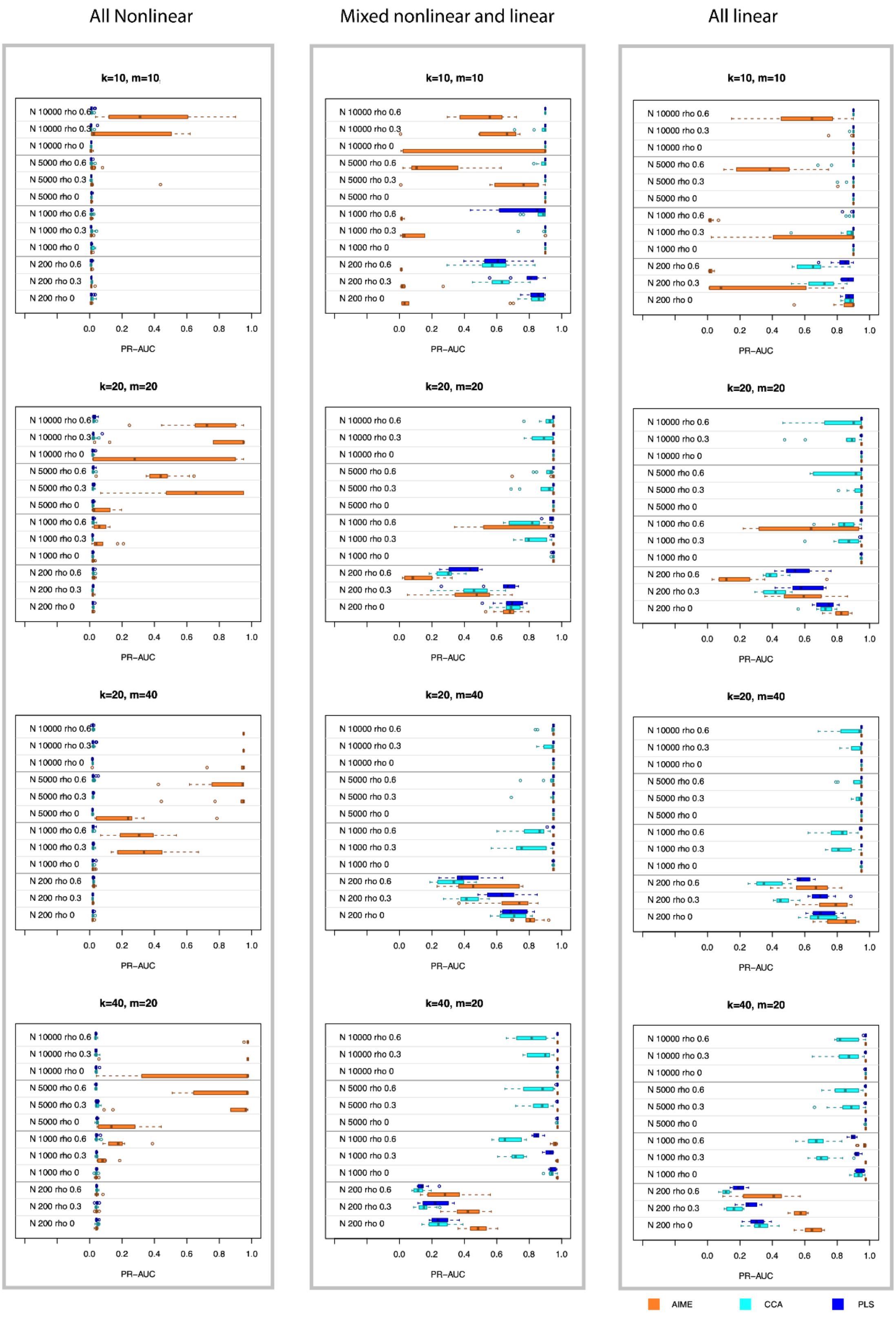

**Figure 2.** Simulation results. PRAUC is used to assess each method's success in selecting the true contributing variables.

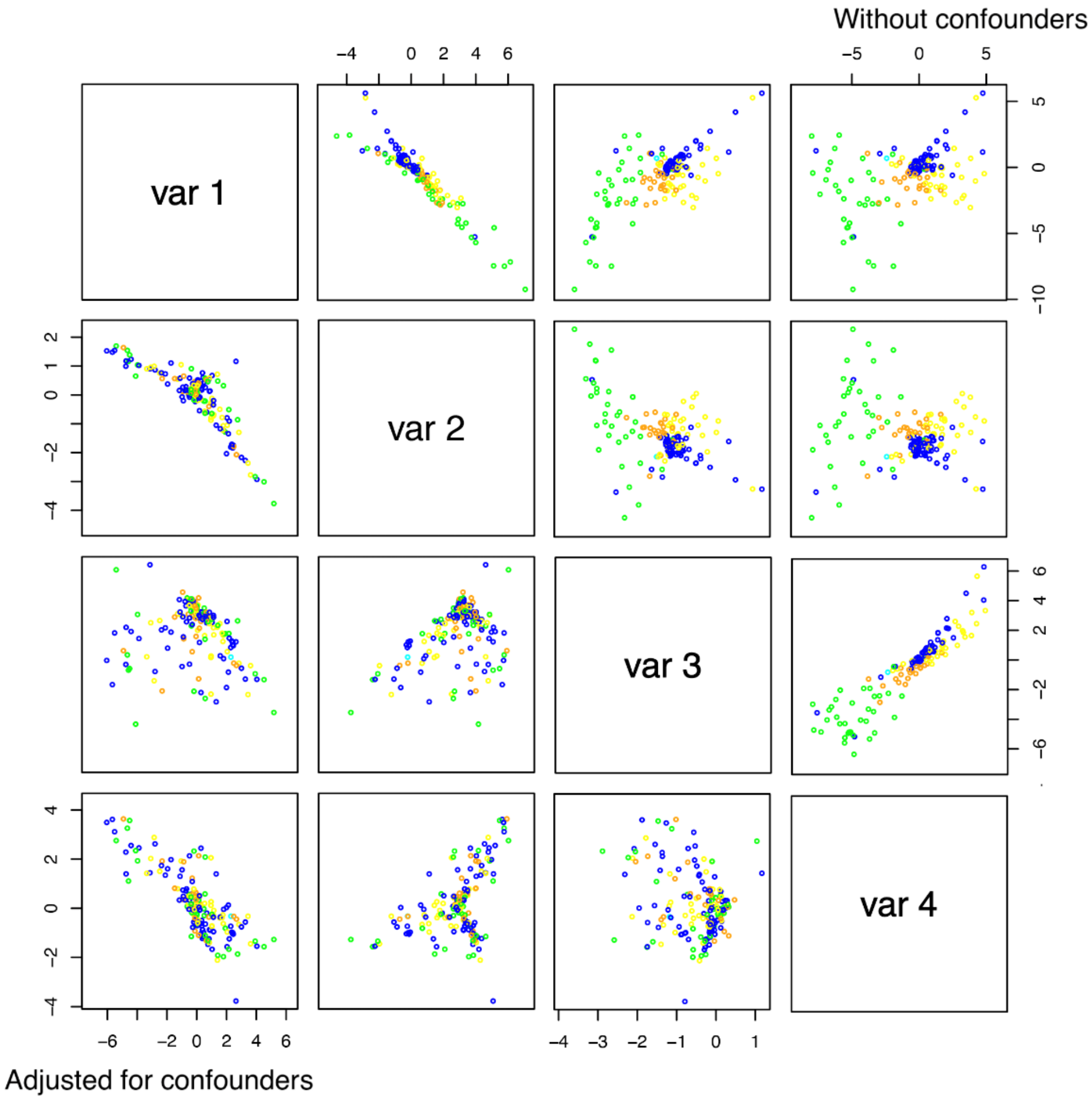

**Figure 3.** AIME results using TCGA miRNA and gene expression data, with and without adjusting for confounders including age, T1 (tumor size) status, and estrogen receptor (ER) status. Points are colored based on PAM50 (Prosigna Breast Cancer Prognostic Gene Signature Assay) subtypes. Upper-right sub-plots: without adjustment for confounders; lower-left sub-plots: with adjustment for confounders.

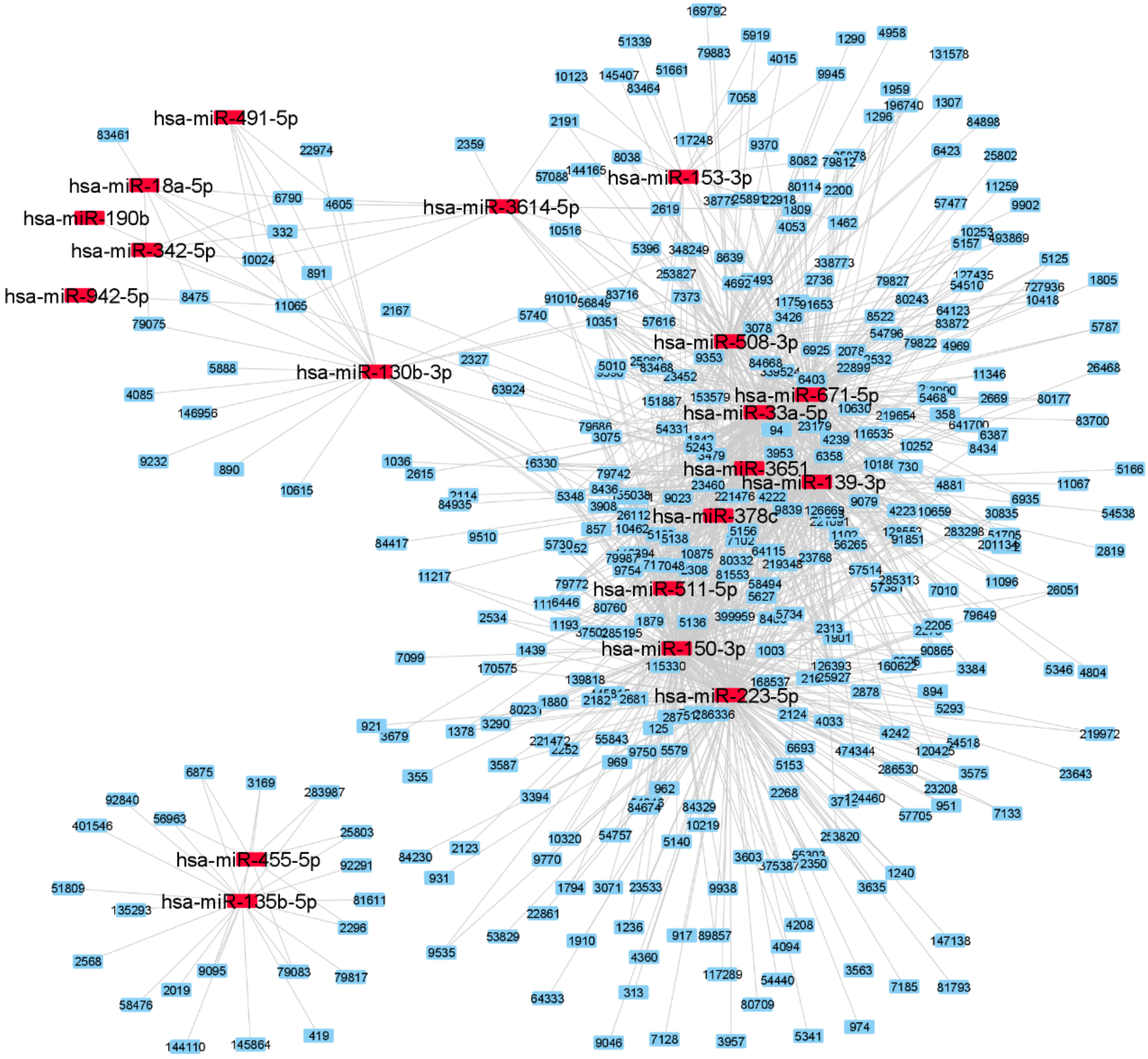

**Figure 4.** The top 20 microRNAs and their major associated genes at the expression level, after adjusting for age, T1 (tumor size) status, and estrogen receptor (ER) status.

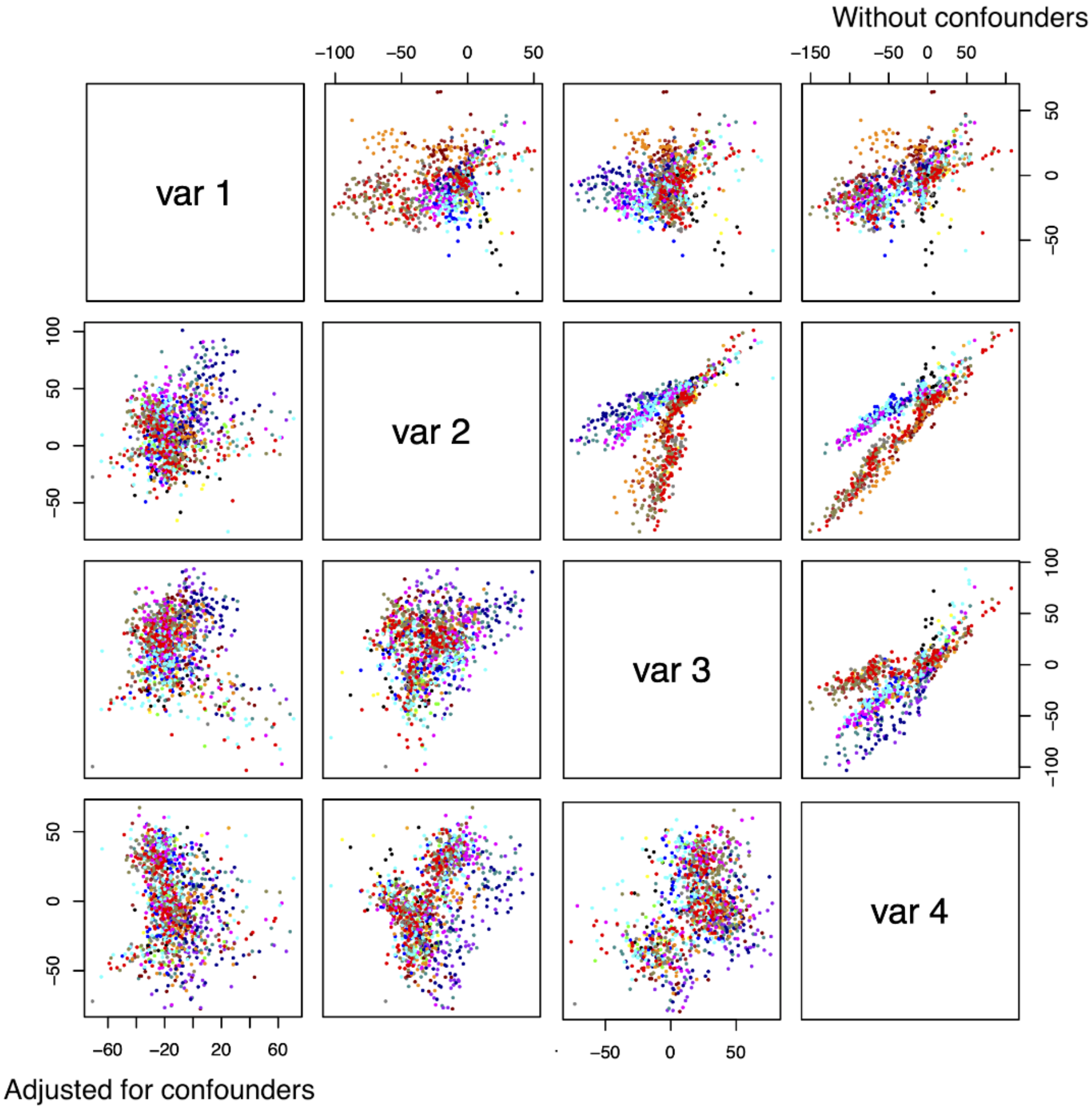

**Figure 5.** AIME results using the MESA Epigenomics and Transcriptomics data (GSE56046/47), with and without adjusting for the “racegendersite” variable of the dataset. DNA methylation M-values were used as input, and the gene expression data as output. Data points are colored based on “racegendersite” variable. Upper-right sub-plots: without adjustment for confounders; lower-left sub-plots: with adjustment for confounders.